\documentclass[aip,final]{revtex4}
\usepackage{amssymb,revsymb}
\usepackage{amsmath}
\usepackage{mathrsfs}
\begin{document}
\title{Phase space gradient of dissipated work and information: 
A role of relative Fisher information}
\author{{\sc Takuya Yamano}}
\email[Email: ]{yamano@amy.hi-ho.ne.jp}
\affiliation{Department of Mathematics and Physics, Faculty of Science, 
Kanagawa University, 2946, 6-233 Tsuchiya, Hiratsuka, Kanagawa 259-1293, Japan}

\begin{abstract}
We show that an information theoretic distance measured by the relative Fisher information 
between canonical equilibrium phase densities corresponding to forward and backward 
processes is intimately related to the gradient of the dissipated work in phase space. 
We present a universal constraint on it via the logarithmic Sobolev inequality. Furthermore, 
we point out that a possible expression of the lower bound indicates a deep connection in 
terms of the relative entropy and the Fisher information of the canonical distributions. 
\medskip
\end{abstract}

\keywords{Relative Fisher information ; Phase space gradient ; 
Logarithmic Sobolev inequality ; Gradient flow}
\pacs{05.20.Gg, 05.70.Ln, 02.30.Sa, 89.70.Cf}

\maketitle
\section{Introduction}
Phase space gradient of thermodynamic quantities in a nonequilibrium process and those 
limitations by information theoretic ones should bring fundamental insights into the 
understandings of the system. Among others, dissipation and work are central quantities 
in thermodynamic operation between equilibrium states. The connecting path between the 
two distinct equilibrium states in phase space has a diversified range depending on 
the steps of the procedure and it determines the amount of the mechanical work needed 
to perform the process. It is well recognized that as long as the process is not 
quasistatic, the free energy at the end of the process becomes less than that of 
the initial state plus the work invested from outside \cite{Landau}. In other words, 
there is dissipative loss of work into the surroundings of the system -- the second 
law of thermodynamics.\\

The dissipated work is thus an indicator of the excess of the injected work into the 
system, and it is defined in terms of the difference in the equilibrium free energy 
$\Delta F$ as $W_{diss}(\Gamma,\lambda)= W(\Gamma,\lambda)-\Delta F$, where $\lambda$ 
is a protocol parameter. $W(\Gamma,\lambda)$ denotes the average work done on the 
system by the external operator as perturbation and it is a function of the position 
in phase space. The dissipated work also represents the total change in entropy 
of the system as a result of the operated transition. The free energy difference 
between terminal states is also called the reversible work. Instead of taking one 
instance in the definition, we consider a statistical ensemble of realizations 
of the process as the consequence of infinitely many repetitions of this process. 
Therefore, we refer to the dissipated work in the sense of mean 
$\langle W_{diss}(\Gamma,\lambda)\rangle$ throughout this paper, but for simplicity 
we omit the angular brackets in the following.\\ 

Any displacement in phase space from a specific equilibrium state to another equilibrium 
phase point induced by a dynamics has a counterpart process by reversing time. The 
to-and-fro movement makes us to expect finding a universal relation in the dissipated 
work in an averaged way. One such intriguing example was discovered as a relation 
\cite{JE2,Kawai}
\begin{eqnarray}
\frac{\langle W_{diss} \rangle}{k T}= D_{KL}(\mathcal{P}_F\|\mathcal{P}_B)\label{eqn:Kawai}, 
\end{eqnarray}
under a sequence of procedure where the system is initially at canonical equilibrium with 
temperature $T$, and then it is detached from the heat reservoir to let the system evolve 
according to the Liouville equation and lastly equilibrate again to a new canonical state 
determined by the protocol parameter by attaching the same bath at the end of the process. 
The equality is replaced by inequality in case the system is kept contacting with the bath 
\cite{Kawai}. $\mathcal{P}_F$ and $\mathcal{P}_B$ are phase densities of the forward and 
the time-reversed processes at a particular time, respectively. Whether or not the prompt 
thermalization for each instance at any attained phase space point is realized remains a 
matter of careful thought, however, it is a standard device for consideration commonly 
used in the literature. The relation gives a meaning of a degree of time-reversal asymmetry 
measured by the Kullback-Leibler (KL) relative entropy $D_{KL}(\mathcal{P}_F\|\mathcal{P}_B)$ 
from $\mathcal{P}_F$ and $\mathcal{P}_B$ \cite{KL,KL2}. The left-hand side of Eq.(\ref{eqn:Kawai}) 
is a physical content in units of thermal energy with the Boltzmann constant $k$ and, on 
the other hand, the right-hand side represents purely information theoretic quantity.\\

In this paper, based on the above consideration, we see how the gradient of the dissipated 
work in phase space is inextricably linked to an information theoretic distance between 
phase density functions of forward and backward processes. Specifically, we show that 
a novel constraint on the averaged square of the gradient of the dissipated work 
$\langle |\nabla W_{diss}|^2\rangle$ can be obtained via the relative Fisher information 
and the logarithmic Sobolev inequality. We further present a possible inequality expression 
indicating a deep connection in terms of two kinds of information quantities. These studies 
are in conformity with a modern approach that bases the construction of statistical 
thermodynamics upon the concept of information (e.g., \cite{Ben}).\\

In Sec.II, we first present a general relation corresponding to Eq.(\ref{eqn:Kawai}) when 
we employ an alternative to the KL relative entropy in order to recognize a distance 
notion in more general context in terms of information measure. The main ingredient of our 
consideration is introduced in Sec.III. We then show a constraint of the obtained relation 
with the inequality in Sec.IV. We give a concluding summary in Sec. V.
\section{A general relation for dissipated work}
Let $M^n$ be the $n$-dimensional configuration space of the present system. The two 
participating distributions $\mathcal{P}_F$ and $\mathcal{P}_B$ are compared at equal-time 
when calculating the general relative entropy, which is specified by a convex function 
$\chi(\mathcal{P}_F/\mathcal{P}_B)$ with $\chi(1)=0$,
\begin{eqnarray}
D_{G}(\mathcal{P}_F\|\mathcal{P}_B):=\int \mathcal{P}_F 
\chi\left(\frac{\mathcal{P}_F}{\mathcal{P}_B}\right)d\Gamma.\label{eqn:DG}
\end{eqnarray} 
This form of the general relative entropy is first introduced in \cite{Csi,Morimoto} and it is  
now well recognized as the Csisz{\'a}r-Morimoto divergence (less known in physics, though). 
It was employed in the proof of the H-theorem for Markov processes \cite{Morimoto}. 
In general, we do not require the asymmetry in the arguments but assume positivity. 
Let $T^*M^n$ be the cotangent bundle to $M^n$, i.e., phase space. The system whose Hamiltonian 
is parameter dependent such as $H_{\lambda}(\Gamma_\lambda)$ starts its evolution from a phase 
space point $\Gamma_0\in T^*M^n$ at time $t_0$ and reaches another point $\Gamma_1\in T^*M^n$ 
at $t_1$. During this interval of time, the protocol parameter changes from 
$\lambda(t_0)=\lambda_0$ to $\lambda(t_1)=\lambda_1$. Initially, the system is put in the 
canonical equilibrium form $\mathcal{P}_F(\Gamma_0)=e^{-\beta H_{\lambda_0}(\Gamma_0)}/Z_0$ 
with $\beta=(k T)^{-1}$. Similarly, prior to the backward process the system is supposed to 
have a density function $\mathcal{P}_B(\Gamma_1)=e^{-\beta H_{\lambda_1}(\Gamma_1)}/Z_1$ by 
preparing the canonical form at the start of the return. The reverse protocol that goes from 
$\lambda_0$ to $\lambda_1$ corresponds to $\Gamma_1$ and $\Gamma_0$ in the phase space. 
The free energy difference between the initial and the final states is given by 
$\Delta F=-\beta^{-1}(\ln Z_1-\ln Z_0)$ with the partition functions $Z_0$ and $Z_1$ for 
each state. We note that the protocol path in $T^*M^n$ joining $\Gamma_0$ to $\Gamma_1$ that 
represents the operation does not necessarily proceed along the geodesic associated with it. 
The geodesic on the surface of the phase space manifold is a significant locus, however, 
the protocol given by an external agent or by an arbitrary schedule does not always go through 
the shortest path. Moreover, even if an operation for one realization takes the geodesic, 
other realizations do not trace exactly the same route at every time. 
They fluctuate around the geodesic path. That is why the work performed on the system is 
statistically distributed representing the ensemble of realizations and it is averaged over 
the ensemble. \\

Let $C$ be a curve (path) on a surface of $M^n$ connecting $\Gamma_0$ and $\Gamma_1$. Then, 
the amount of work the system receives on completion of the process depends not only on the 
location of the two terminal points $\Gamma_0$ and $\Gamma_1$, but also on the route $C$. 
It should be precisely denoted as $W^C(\Gamma_0,\Gamma_1)$ and equals to the difference in 
the Hamiltonian $H_{\lambda_1}(\Gamma_1) - H_{\lambda_0}(\Gamma_0)=\int^{\lambda_1}_{\lambda_0}d\lambda 
\partial H_\lambda(\Gamma_\lambda)/\partial \lambda$ (the first law of thermodynamics). 
For simplicity, we omit both the subscript and arguments, and we denote $W$ in the following.\\

We note that for the well-defined distance $D_G$, the relative density $\mathcal{P}_F/\mathcal{P}_B$ 
must have a finite value on a support (that is, $\mathcal{P}_F$ is absolutely continuous 
with respect to $\mathcal{P}_B$), otherwise it is defined as infinity. Then, the relative 
phase density becomes 
\begin{eqnarray}
\frac{\mathcal{P}_F}{\mathcal{P}_B}=\frac{Z_1}{Z_0}
e^{-\beta\left(H_{\lambda_0}(\Gamma_0) - H_{\lambda_1}(\Gamma_1)\right)}
= e^{\beta (W-\Delta F)}.\label{eqn:reld}
\end{eqnarray}
Inserting this into Eq.(\ref{eqn:DG}), we obtain the general reformulation of 
Eq.(\ref{eqn:Kawai}) as the following form 
\begin{eqnarray}
D_{G}(\mathcal{P}_F\|\mathcal{P}_B)=\Big\langle\chi(e^{\beta(W-\Delta F)})
\Big\rangle_{\mathcal{P}_F},
\end{eqnarray}
where $\langle \cdot \rangle_{\mathcal{P}_F}$ denotes the average with respect to the 
distribution $\mathcal{P}_F$. This is the most general expression relating the dissipated work 
and the distance measure in this setting. We remark that the choice $\chi(x)=\ln x$ in 
Eq.(\ref{eqn:DG}) with $x=\mathcal{P}_F/\mathcal{P}_B$ defines the KL distance and recovers 
exactly the result Eq.(\ref{eqn:Kawai}). It is also worth mentioning another example. If we choose 
$\chi(x)=1/\sqrt{x}$, we have the overlap distance between $\mathcal{P}_F$ and $\mathcal{P}_B$, 
i.e., $D_O(\mathcal{P}_F\|\mathcal{P}_B):=\int \sqrt{\mathcal{P}_F\mathcal{P}_B} d\Gamma
= \langle \sqrt{\mathcal{P}_F/\mathcal{P}_B}\rangle_{\mathcal{P}_F}$. 
If two phase densities are identical, it gives unity. However, $\mathcal{P}_F$ and 
$\mathcal{P}_B$ can never be identical ($\mathcal{P}_F\neq \mathcal{P}_B$) in our consideration. 
Substituting Eq.(\ref{eqn:reld}) stemming from the canonical forms into the definition of the 
overlap distance, we have 
\begin{eqnarray}
D_O(\mathcal{P}_F\|\mathcal{P}_B)&=&
\Big\langle e^{-\frac{\beta}{2}(W-\Delta F)} \Big\rangle_{\mathcal{P}_F}\nonumber\\
&=& e^{\frac{\beta}{2}\Delta F}\Big\langle e^{-\frac{\beta}{2}W}\Big\rangle_{\mathcal{P}_F}\nonumber\\
&\stackrel{\text{J.E.}}{=}& 0. \label{eqn:DO}
\end{eqnarray}
The second line follows from the fact that the $\Delta F$ is not a statistically distributed quantity.
The last line is due to the nonequilibrium work relation called the Jarzynski equality (J.E.) 
$\langle e^{-\beta W}\rangle=e^{-\beta \Delta F}$ \cite{JE1,Crooks}, 
where the angular brackets denotes an average over a statistical ensemble of realizations of a 
single process through phase space. It is known that for general circumstances including the 
Hamilton's evolution, this average is equivalent to taking an average using the initial 
equilibrium distribution designated by $\lambda_0$ \cite{JE3}. Furthermore, the J.E. can be 
potentially useful in various equilibrium statistical physics, in which the estimation of the 
free energy differences is difficult to obtain. In these cases, such an equilibrium property 
is obtained from observations (experiments) of work distributions performed on the system.
Note that distinct distributions in phase space do not share the common domain, so that 
Eq.(\ref{eqn:DO}) means that the vanishing overlap between the two equilibrium states is consistent 
with the fact that the J.E. holds. On the other hand, the J.E. is irrelevant to the derivation of 
Eq.(\ref{eqn:Kawai}), i.e. the case of the KL distance. We also remark that when the system realizes 
$W=\Delta F$ (reversible process), the overlap $D_o$ vanishes, meaning that it can be a measure of 
how much the nonequilibrium behavior deviates from the reversibility. 
\section{Relative Fisher information and dissipated work}
In Section II, we recognized that Eq.(\ref{eqn:Kawai}) is just one realization  
of the information theoretic expression of the averaged dissipated work among other possibilities. 
The choice of $\chi$ is basically at our disposal and there is no legitimate criterion in terms 
of information theory. For operational purposes such as statistical inferences and estimation 
of measurement values, relative entropy with parameters can be considered advantageous over 
the one without them. Such a generalized distance measure with keeping the original properties 
of the Kullback-Leibler has been proposed \cite{TY2} (see also e.g., \cite{Special}). But we 
do not delve into such possibilities including the overlap distance in this paper. Instead, 
to advance the understanding in line with previous works, we hereafter set the gauge of the 
distance measure as KL in our consideration. As we pronounced in Introduction, we are concerned 
with the quantity $\nabla_{\Gamma} W_{diss}$, where $\nabla_{\Gamma}$ indicates the gradient 
in phase space, because the dissipated work depends on the location in phase space and its 
gradient reflects a local structure (geometry) of the dissipation occurred as an excess of 
work performed on the system. \\

The relative Fisher information from $f$ to $g$ is defined as (e.g. \cite{Villani}, p.278)
\begin{eqnarray}
D_{\rm RFI}(f\|g):=\int f \Big| \nabla_\Gamma \left(\ln \frac{f}{g}\right)\Big|^2 d\Gamma.
\label{eqn:rFi}
\end{eqnarray}
It is non-negative and achieves zero iff $f=g$. Further, it is asymmetric, i.e., 
$D_{\rm RFI}(f\|g)\neq D_{\rm RFI}(g\|f)$, meaning that it is {\it directed} as is the case of 
the KL distance. This form can be defined independently of whether the distributions possess 
estimation parameters or not. We can also employ it when the two distributions are parametrized 
by the same family (say, $\boldsymbol{\theta}$). We can say that this metric reflects a local 
comparison of the two distributions in that we take the derivative of them, whereas the KL 
distance returns a coarse-grained quantity by gathering each contribution of the displacement 
(or difference) between distributions. The same remark mentioned before Eq.(\ref{eqn:reld}) 
also applies to this relative distance: the relative phase density must have finite value on 
the domain of phase space. To the best of our knowledge, there is no study to give a physical 
connection to that measure. Now, we show one of the applications below. That is, substituting 
Eq.(\ref{eqn:reld}) into the above definition, we have a relation (see Appendix \ref{app:rfi})
\begin{eqnarray}
D_{\rm RFI}(\mathcal{P}_F\|\mathcal{P}_B)=\beta^2\Big\langle \Big| \nabla_\Gamma 
W_{diss}(\Gamma)\Big|^2 \Big\rangle_{\mathcal{P}_F}\label{eqn:reF1}.
\end{eqnarray}
This identity articulates that an average of the square of the gradient of the dissipated work 
generated by the transient process, taken over a forward equilibrium state, can be equated with 
the distance between the forward and backward distributions in the phase space measured by the 
relative Fisher information.\\

Our setting of formulation represented by an operational parameter is in a position of 
Hamilton equations (differential equations on symplectic manifolds). Then, in order 
to consider nonequilibrium processes in phase space perspective, it is principally 
necessary to deal with the trajectory on a manifold (either on symplectic or on Riemannian). 
It is known that paracompact $C^\infty$-manifolds always have their Riemannian metrics. 
Since the usual manifold is paracompact, the Riemannian metric is introduced if we 
regard the phase space as a manifold. Then, the metric expression of the above identity 
is of concern. To this end, we remark that a gradient vector of a differentiable function 
$f$ on the Riemannian manifold associated with the covector $df$ can be defined as the 
contravariant vector, which is given in coordinate as 
$(\nabla f)^i=\sum_j g^{ij}\partial f/\partial x^j$, where $x^i$ represents the generalized 
coordinates and the contravariant metric tensor $g^{ij}$ is the inverse of the metric tensor 
$g_{ij}$ ($(g^{ij})=(g_{ij})^{-1}$). We further recall that, for any vector ${\bf a}$, the 
differential of a function $df$ is defined by the derivative of $f$ along ${\bf a}$.   
Therefore, the gradient vector $\nabla f$ satisfies 
$df({\bf a})=\langle \nabla f, {\bf a} \rangle=\sum_i (\partial f/\partial x^i) a^i$.  
Then, substituting $(\nabla f)^i$ into this, we find that the square of the modulus of the 
gradient $f$ can be written as $|\nabla f|^2=df(\nabla f)=\sum_{ij}(\partial_{x^i}f)g^{ij}(\partial_{x^j}f)$.
Since we now take $f=W_{diss}(\Gamma)$ in the present consideration, therefore, the right-hand 
side of Eq.(\ref{eqn:reF1}) can be expressed as 
\begin{eqnarray}
\beta^2 \Big\langle \sum_{i j}\frac{\partial W_{diss}(\Gamma)}{\partial x^i}g^{ij}
\frac{\partial W_{diss}(\Gamma)}{\partial x^j} \Big\rangle_{\mathcal{P}_F}.
\end{eqnarray}
We can obtain a more direct physical meaning for the quantity of the Dirichlet form 
$\langle | \nabla_\Gamma W_{diss}(\Gamma)|^2 \rangle_{\mathcal{P}_F}$ appearing in 
Eq.(\ref{eqn:reF1}) via the gradient flow interpretation as we will see shortly. To this end, we 
recall again what $W$ exactly expresses. It means that the work done on the system when it starts 
from $\Gamma_0$ and reaches $\Gamma_1$ by changing the control parameter $\lambda\in [\lambda_0,\lambda_1]$ 
along a path $C$ on the surface of the phase space manifold. Since $\lambda$ designates the instance 
of the evolution from any starting point $\Gamma_0$, we rewrite it as $W_\lambda(\Gamma)$ by denoting 
$\Gamma_0$ as $\Gamma$ on a specific $C$. Accordingly, the dissipated work up to the intermediate value 
of $\lambda$ can be denoted as $W_{diss}(\Gamma_\lambda)$, where the phase point $\Gamma_\lambda\in T^*M^n$ 
should be read as $\hat{T}_\lambda\Gamma$ with an evolution operator $\hat{T}_\lambda$ that represents 
the protocol. Recall now that the gradient flow associated with the velocity vector field 
$-\nabla(\delta G/\delta \rho):=v(\Gamma_\lambda)$ is defined (Appendix \ref{app:gradf}) as 
\begin{eqnarray}
\frac{\partial \rho}{\partial t}={\rm div} \left(\rho \nabla\frac{\delta G}{\delta \rho}\right),\label{eqn:gradf}
\end{eqnarray}
where $G(\rho)$ is an energy functional. In our case, we can take it as 
\begin{eqnarray}
G(\rho)=\int W_{diss}(\Gamma_\lambda) \rho(\Gamma)d\Gamma.
\end{eqnarray} 
With abuse of notation above, we have used $\rho(\Gamma)$ as the probability density 
function in phase space at each time instant $t$. This is the phase space average of the dissipated work 
in our consideration and the first variation of it is $\delta G/\delta \rho= W_{diss}(\Gamma_\lambda)$.  

Therefore, the average kinetic energy $K_\lambda$ accompanied to the dissipation up to the protocol 
$\lambda$ is found to be just the squared mean of the gradient of the dissipated work evaluated by 
the initial equilibrium state $\mathcal{P}_F$ that is prepared for the start of the process: 
\begin{eqnarray}
K_\lambda=\int |v(\Gamma_\lambda)|^2\mathcal{P}_F(\Gamma)d\Gamma 
= \Big\langle | \nabla_\Gamma W_{diss}(\Gamma)|^2 \Big\rangle_{\mathcal{P}_F}.
\end{eqnarray}
In the whole process, the total kinetic energy becomes $\int^{\lambda_1}_{\lambda_0}K_\lambda d\lambda$. 
\section{A lower bound for average dissipated-work gradient}
In this section, we show that our primarily focused quantity can be bounded by the KL distance 
via the logarithmic Sobolev inequality (LSI) \cite{Gross}. In this sense, the identity that 
connects the relative Fisher information and the quantity 
$\langle | \nabla_\Gamma W_{diss}(\Gamma)|^2 \rangle_{\mathcal{P}_F}$ derived in the previous 
section can be regarded as an intermediate result. The LSI has wide range of applications and it 
takes several mathematically equivalent forms \cite{Villani}. In this paper, we employ the form 
(Lemma 6.1 in \cite{Gross})
\begin{eqnarray}
\int |f|^2 \ln |f|d\mu \leqslant c \int |\nabla f|^2d\mu + \|f\|_2^2\ln \|f\|_2^2,\label{eqn:LSI}
\end{eqnarray}
which holds for all functions $f$, whose gradient $\nabla f$ and $f$ itself are square 
integrable in the domain. $\|f\|_2$ denotes $(\int f^2 d\mu )^{1/2}$ and the constant $c>0$ 
independent of $f$. The physical interpretation of this constant is not obvious in general 
circumstances and would depend on the physical model. However, when an equilibrium state 
satisfies the LSI and the probability density function has the corresponding gradient flow, 
the constant has a clear meaning of how fast the system equilibrates with it. The corresponding 
relaxation rate takes the exponential form and it appears as a pre-factor to the KL distance  
between the initial and the equilibrium density functions. This feature is a direct consequence 
of the result provided in \cite{Barron} (see also e.g., \cite{Villani}, p.288 and Appendix 
\ref{app:LSI}). If we substitute $d\mu=\mathcal{P}_B(\Gamma)d\Gamma$ for the probability measure 
and choosing $f=\sqrt{\mathcal{P}_F/\mathcal{P}_B}$ then multiplying both sides by $2$, we find  
\begin{eqnarray}
\int \mathcal{P}_F \ln \frac{\mathcal{P}_F}{\mathcal{P}_B}d\Gamma & \leqslant & 
2c\int \Big| \nabla \sqrt{\frac{\mathcal{P}_F}{\mathcal{P}_B}}\Big|^2 \mathcal{P}_B d\Gamma \nonumber\\
& = & \frac{c}{2} \int \mathcal{P}_F \Big| \nabla \left(\ln \frac{\mathcal{P}_F}
{\mathcal{P}_B}\right)\Big|^2d\Gamma.\label{eqn:LSI2}
\end{eqnarray}
In what follows, we set $c=1$ without impeding our consideration. Combining Eq.(\ref{eqn:reF1}) 
and Eq.(\ref{eqn:LSI2}), we find that the mean of the gradient dissipated work is lower bounded 
by the information theoretic distance between forward and backward phase densities :
\begin{eqnarray}
\langle | \nabla_\Gamma W_{diss}(\Gamma)|^2 \rangle_{\mathcal{P}_F} \geqslant 
2(kT)^2 D_{KL}(\mathcal{P}_F \| \mathcal{P}_B).\label{eqn:NW}
\end{eqnarray}
We further pursue a reformulation of thus obtained constraint from an information point of view, 
that is, in terms of Fisher information in statistics. \\

We begin with recalling the followings. For a family of probability distributions 
$\rho_{\boldsymbol{\theta}}(\boldsymbol{x})$ parametrized by $\boldsymbol{\theta}$, the Fisher 
information in estimation theory is defined as 
$I(\boldsymbol{\theta}):=\langle (\nabla_{\boldsymbol{\theta}} \rho/\rho)^2\rangle_{{\rho}_
{\boldsymbol{\theta}}(\boldsymbol{x})}$. In the same way, it is defined also as 
$I(\rho(\boldsymbol{x})):=\langle (\nabla_{\boldsymbol{x}} \rho/\rho)^2\rangle_{{\rho}(\boldsymbol{x})}$ 
for a differentiable distribution $\rho(\boldsymbol{x})$, and it measures how much two neighboring 
density functions are statistically distinguishable \cite{Cover}. The former form is invariant 
against any shift $\boldsymbol{\theta}$, which signifies the diagonal entries of the Fisher 
information matrix, thereby implying i.i.d. data. Indeed, when we choose as 
$\rho_{\boldsymbol{\theta}}(\boldsymbol{x})=\rho(\boldsymbol{x}-\boldsymbol{\theta})$, 
we easily find that due to 
$\nabla_{\boldsymbol{x}}\rho(\boldsymbol{x}-\boldsymbol{\theta})=-\nabla_{\boldsymbol{\theta}}
\rho(\boldsymbol{x}-\boldsymbol{\theta})$, the Fisher information does not depend on 
$\boldsymbol{\theta}$ and $I(\boldsymbol{\theta})=I(\rho(\boldsymbol{x}))$ follows.\\

There have been comprehensive efforts to understand various physical laws in terms of this 
information \cite{Frieden}. It also plays a crucial role to upper bound the entropy production 
(e.g. \cite{TY1} and references therein). Since the present canonical equilibrium distribution 
specified by a protocol parameter satisfies $\nabla\mathcal{P}(\Gamma)=-\beta H(\Gamma)\mathcal{P}(\Gamma)$, 
it becomes  
\begin{eqnarray}
I(\mathcal{P}) = \Big\langle \left(\frac{\nabla \mathcal{P}}{\mathcal{P}}\right)^2\Big\rangle_{\mathcal{P}}
= \int \frac{\nabla \mathcal{P}}{\mathcal{P}} (-\beta \nabla H(\Gamma)) \mathcal{P}d\Gamma.\nonumber\\
\end{eqnarray}
Integrating by parts, the right-hand side reduces further to 
\begin{eqnarray}
-\beta \int \nabla H(\Gamma) \nabla \mathcal{P}(\Gamma)d\Gamma 
=\beta \langle \Delta H(\Gamma) \rangle_{\mathcal{P}},\nonumber
\end{eqnarray}
where we have assumed that the phase density vanishes at the boundary ($\partial S$) of the 
system, so that $[\mathcal{P}\nabla H(\Gamma)]_{\partial S}=0$, and $\Delta$ denotes the 
Laplacian in phase space. Therefore, we obtain a striking relation
\begin{eqnarray}
I(\mathcal{P}) &=& \beta \langle \Delta H(\Gamma) \rangle_{\mathcal{P}},\label{eqn:Ftmp}
\end{eqnarray}
which tells that the inverse temperature of the system can be intimately linked with the 
information quantity (Upon revision of this manuscript, the author became aware 
that this relation has derived also in Ref. \cite{Narayanan}.). 
In general coordinates, we note that $\Delta H(\Gamma)$ can be expressed as 
$(\sqrt{g})^{-1}\partial_i(\sqrt{g}g^{ij}\partial_jH(\Gamma))$, where $g$ is the determinant 
of the metric $g=det(g_{ij})$. A way to interpret this relation is that the temperature of 
the system can be defined by the Fisher information of the system and by the averaged second 
order differential (curvature) of energy \cite{Ffl}. As we shall use below, this relation 
is critical for the present study. Since the system contacts with the same heat bath (with 
common $\beta$) both at the start and the end of the process as described in Sec. II 
(or we could also restate it as follows; the heat reservoir is so large compared with the 
system, so that temperature of the reservoir is not disturbed by the heat discarded by 
the system), we have readily a relation from Eq.(\ref{eqn:Ftmp}) 
\begin{eqnarray}
\frac{I(\mathcal{P}_F)}{\langle \Delta H(\Gamma) \rangle_{\mathcal{P}_F}}= 
\frac{I(\mathcal{P}_B)}{\langle \Delta H(\Gamma) \rangle_{\mathcal{P}_B}}=
\beta.\label{eqn:Iofb}
\end{eqnarray}
An immediate but profound implication of this consequence is that the ratio of the Fisher 
information associated with probability distributions of the forward and backward processes 
is equivalent to the ratio of the averaged curvature of the Hamiltonian. This is true   
if the system takes the canonical form in distribution. The former ratio has the origin of the 
information quantity and the latter has the physical one. Next, substituting Eq.(\ref{eqn:Iofb}) 
into Eq.(\ref{eqn:NW}), we readily have an inequality
\begin{eqnarray}
\frac{\langle | \nabla_\Gamma W_{diss}(\Gamma)|^2 \rangle_{\mathcal{P}_F}}
{\langle \Delta H(\Gamma) \rangle_{\mathcal{P}_{\gamma}}^2} \geqslant 
2\frac{D_{KL}(\mathcal{P}_F \| \mathcal{P}_B)}{[I(\mathcal{P}_\gamma)]^2}.\label{eqn:last}
\end{eqnarray}
where the symbol $\gamma$ denotes either $F$ or $B$ representing the forward and the backward 
equilibrium states. The lower bound on the ratio relevant to the averaged physical quantities 
(the left-hand side) is nicely bounded from below in terms of the ratio of information-associated 
quantities only. 
A further interesting observation can be derived for this relation Eq.(\ref{eqn:last}) from  
the well-known Cramer-Rao inequality in statistical estimation theory \cite{Cover}. 
For simplicity's sake, we consider it for one-dimensional case. The Cramer-Rao inequality tells 
a tradeoff relation between Fisher information of a distribution $P(X)$ and the variance of 
the distribution $\sigma^2_X$, i.e., $I(P(X))\geqslant 1/\sigma^2_X$, where $X$ is a random 
variable. Then, we find that the lower bound in Eq.(\ref{eqn:last}) is upper bounded by 
$2D_{KL}(\mathcal{P}_F \| \mathcal{P}_B) \sigma^4_X$. Now that $P(X)$ is of the canonical 
form in our setting, the equality can be achieved when the Hamiltonian $H(X)$ is of 
quadratic form. 
\section{Summary} 
To deepen the understandings of a profound information theoretic relation between work and 
dissipation in nonequilibrium systems, we have derived a universal relation that connect the 
gradient of the dissipated work and the relative Fisher information within a framework of 
the setup repeatedly employed in the previous studies. Considering the gradient of the 
dissipated work at each point in phase space enables us to get geometric information that 
cannot be obtained from KL entropy only. The relative Fisher information plays the role. 
By way of this, we have established the information based lower bound for the quantity 
relevant to the dissipation in phase space. The instantaneous equilibration was a premise to 
assure the well-posedness of the free energy difference between the two canonical equilibrium 
states associated to the forward and backward processes. However, the notion of the 
nonequilibrium free energy change $\Delta F_{neq}$ has recently considered to refine the 
dissipation occurring in far from thermodynamic equilibrium, where the dissipated work 
is defined by average work minus $\Delta F_{neq}$ instead of the equilibrium free energy 
change $\Delta F$ \cite{Fneq}. An extension of the present result to such a case surely 
needed if one is to understand and to gain deeper insights into biological systems.
\begin{acknowledgments}
The author wishes to thank Hiroaki Yoshida for a valuable discussion on the relative Fisher 
information at the Ochanomizu University in August 2012.
\end{acknowledgments}
\appendix
\section{Relative Fisher information $D_{\rm{RFI}}$}\label{app:rfi}
The concept of the relative Fisher information measured from $f$ to $g$, $D_{\rm{RFI}}(f\|g)$ 
is less acknowledged in application by the general physics community compared with the relative 
entropy (Kullback-Leibler divergence). We briefly present here a physical origin of this form 
that is unrecognized in the literature (see also \cite{TY3} and references therein for a more 
general discussion). We consider the time change of the KL divergence between two distribution 
functions $f=f(\boldsymbol{x},t)$ and $g=g(\boldsymbol{x},t)$ that obey a heat equation 
($\partial_tf=\mathscr{D}\nabla^2f$ etc.), where $\mathscr{D}$ is a diffusion constant:
\begin{eqnarray}
\frac{d}{dt} D_{KL}(f\|g)=\frac{d}{dt}\int f\ln \frac{f}{g}d\boldsymbol{x} =
\int \dot{f} \ln \frac{f}{g}d\boldsymbol{x} + 
\int g\left(\frac{\dot{f}}{g}-\frac{f}{g^2}\dot{g} \right) d\boldsymbol{x},
\end{eqnarray}
where $\dot{f}=\partial f/\partial t$ etc. The first term is calculated by integration by 
parts as 
\begin{eqnarray}
\mathscr{D}\int (\nabla^2f)\ln \frac{f}{g} d\boldsymbol{x}=-\mathscr{D} \int 
(\nabla f)\nabla \left( \ln \frac{f}{g}\right)d\boldsymbol{x},\label{eqn:1stt}
\end{eqnarray}
where we have used the boundary conditions that $f$ and $\nabla f$ vanish when $|\boldsymbol{x}|\to 0$. 
Similarly, under the conditions that $f$, $g$ and $\nabla g$ vanish when $|\boldsymbol{x}|\to 0$, 
the second term becomes 
\begin{eqnarray}
\mathscr{D}\int f\left( \frac{\nabla g}{g}\right) \nabla\left( \ln \frac{f}{g}\right) d\boldsymbol{x}.
\label{eqn:2ndt}
\end{eqnarray}
Therefore, combining Eq.(\ref{eqn:1stt}) and Eq.(\ref{eqn:2ndt}), we have 
\begin{eqnarray}
\frac{d}{dt} D_{KL}(f\|g)&=&-\mathscr{D}\int f\left( \frac{\nabla f}{f} - 
\frac{\nabla g}{g}\right) \nabla\left( \ln \frac{f}{g}\right) d\boldsymbol{x}\nonumber\\
&=& -\mathscr{D}\int f \Big|\nabla\left( \ln \frac{f}{g}\right)\Big|^2d\boldsymbol{x}.\label{eqn:dBi}
\end{eqnarray}
Except for the diffusion coefficient $\mathscr{D}$, the right-hand side provides $D_{\rm{RFI}}$, 
which is called the de Bruijn-type identity \cite{TY3}. The information involves how fast 
the KL distance changes under the heat equation. In a more general context \cite{TY3}, we have 
\begin{eqnarray}
\frac{d}{dt} D_{KL}(f\|g)&=& \Big\langle \left( \frac{\boldsymbol{j}_f}{f}-
\frac{\boldsymbol{j}_g}{g}\right)\nabla\left( \ln\frac{f}{g}\right)\Big\rangle_f,\label{eqn:gdBi}
\end{eqnarray}
where $\boldsymbol{j}_f$ and $\boldsymbol{j}_g$ are the flows associated, respectively, with 
$f$ and $g$ in the continuity equation $\partial_t\rho=-\nabla\cdot \boldsymbol{j}$,  
and $\langle \cdot\rangle_f$ denotes the averaging with respect to $f$.
\subsection{$D_{\rm{RFI}}$ between two canonical equilibrium distributions}
The relative Fisher information between forward and backward phase densities is calculated as 
\begin{eqnarray}
D_{\rm{RFI}}(\mathcal{P}_F \| \mathcal{P}_B) &=& \int \mathcal{P}_F
\Big| \nabla_\Gamma \left(\ln \frac{\mathcal{P}_F}{\mathcal{P}_B}\right)\Big|^2d\Gamma 
\quad (\text{by definition Eq.(\ref{eqn:rFi})})\nonumber\\
&=& \int \mathcal{P}_F \Big| \nabla_\Gamma \left(\ln e^{\beta(W-\Delta F)}\right)\Big|^2 d\Gamma 
\quad (\text{by relation Eq.(\ref{eqn:reld})}) \nonumber\\
&=& \beta^2\int \mathcal{P}_F \Big| \nabla_\Gamma (W-\Delta F)\Big|^2 d\Gamma.
\end{eqnarray}
Finally, by the definition of the dissipated work, we have the relation Eq.(\ref{eqn:reF1}).
\section{Gradient flow}\label{app:gradf}
An evolution equation of the form 
\begin{eqnarray}
\frac{\partial \rho}{\partial t}={\rm grad_{W}}G(\rho)
\end{eqnarray}
for some functionals $G(\rho)$, is called a gradient flow, where ${\rm grad_{W}}G(\rho)$ is the 
Wasserstein gradient on the Wasserstein space (e.g., \cite{Villani}). A wide class of partial 
differential equations can be understood in light of this approach. The Wasserstein gradient here 
${\rm div}(\rho v(\Gamma_\lambda))$ for the potential functional $G$ causes the time change in the 
density function. In other words, the vector field makes the backward density function $\mathcal{P}_B$ 
different from the forward one $\mathcal{P}_B$, during which the work dissipation is completed. 
The variation of $G$ with respect to $\rho$ is the gradient in $L^2$, and corresponds to the 
dissipated work $W_{diss}(\Gamma_\lambda)$ in the present consideration. To be more convinced, 
it is appropriate to consider a linear Fokker-Planck equation whose free energy functional $G(\rho)$ 
can be given by 
\begin{eqnarray}
G(\rho) = \int u(\boldsymbol{x})\rho(\boldsymbol{x})d\boldsymbol{x} + 
T\int \rho(\boldsymbol{x})\ln\rho(\boldsymbol{x})d\boldsymbol{x},
\end{eqnarray}
with $u(\boldsymbol{x})$ and $T$ denoting the internal energy and temperature, respectively. Then, 
the gradient of the functional $\delta G/\delta \rho=u(\boldsymbol{x})+\log\rho(\boldsymbol{x})+1$ 
leads to the gradient flow $\partial \rho/\partial t={\rm div}(\nabla\rho+\rho\nabla u(\boldsymbol{x}))$. 
In terms of the continuity equation, it is equivalent to that the system has a flow 
$\boldsymbol{j}=-(\nabla \rho+\rho\nabla u(\boldsymbol{x}))$. That is, it consists of the Fick's 
law plus the gradient of the internal energy. Substituting this into Eq.(\ref{eqn:gdBi}), 
we immediately recover the de Bruijn-type identity Eq.(\ref{eqn:dBi}).
\section{Approach to equilibrium controlled by a constant in the logarithmic Sobolev inequality}
\label{app:LSI}
The constant $c$ in the LSI (Eq.(\ref{eqn:LSI}): 
$\int|f|^2\ln|f|d\mu-(\int |f|^2d\mu)\ln(\int |f|^2d\mu)\leqslant c\int |\nabla f|^2d\mu$) 
determines the rate at which the system approaches to an equilibrium state when measured with the KL 
distance. Consider the KL distance between a density function $(f_t)_{t\geqslant 0}$ and an 
equilibrium one $f_\infty$. Its time derivative is then 
\begin{eqnarray}
\frac{d}{dt}D_{KL}(f_t\|f_\infty)=\int \dot{f}_t\ln \frac{f_t}{f_\infty}d\boldsymbol{x}.
\end{eqnarray}
Since we take $G=D_{KL}(f_t\|f_\infty)$ in the definition (Eq.(\ref{eqn:gradf})) and then 
$\delta G/\delta f_t=1+\log f_t/f_\infty$, the gradient flow associated with it is 
\begin{eqnarray}
\frac{\partial f_t}{\partial t}=\nabla\cdot\left(f_t\nabla\ln\frac{f_t}{f_\infty}\right).
\end{eqnarray}
Substituting this into the above and performing the integration by parts under the vanishing 
boundary condition, we have 
\begin{eqnarray}
\frac{d}{dt}D_{KL}(f_t\|f_\infty)=-\int f_t\Big|\nabla \ln\frac{f_t}{f_\infty}\Big|^2d\boldsymbol{x}
=-D_{\rm RFI}(f_t\|f_\infty).\label{eqn:C3}
\end{eqnarray}
This is the de Bruijn-type identity under a gradient flow with the equilibrium state $f_\infty$ as 
a reference density function. On the other hand, choosing the probability measure as 
$d\mu=f_\infty d\boldsymbol{x}$ and putting $f=\sqrt{f_t/f_\infty}$ in the LSI, we have 
\begin{eqnarray}
\frac{1}{2}\int \frac{f_t}{f_\infty}\ln\left(\frac{f_t}{f_\infty}\right)f_\infty d\boldsymbol{x} 
& \leqslant & c\int \Big| \nabla\sqrt{\frac{f_t}{f_\infty}}\Big|^2f_\infty d\boldsymbol{x}\nonumber\\
&=&\frac{c}{4}\int f_t\Big| \nabla \ln \frac{f_t}{f_\infty}\Big|^2f_\infty d\boldsymbol{x}\nonumber.
\end{eqnarray}
The right-hand side equals just to $c/4 D_{\rm{RFI}}(\mathcal{P}_F \| \mathcal{P}_B)$. 
Therefore, from  Eq.(\ref{eqn:C3}) we have an inequality 
\begin{eqnarray}
\frac{d}{dt}D_{KL}(f_t\|f_\infty) \leqslant -\frac{2}{c} D_{KL}(f_t\|f_\infty).\label{eqn:C4}
\end{eqnarray}
Recall that the Gronwall's inequality states that if two continuous functions $w(t)$ and $\gamma(t)$ 
defined on an interval $I=[a,\infty)$ satisfies a differential inequality 
$w^\prime(t)\leqslant \gamma(t)w(t)$, then $w(t)$ is bounded as 
$w(t)\leqslant w(a)\exp(\int^{t}_{a}\gamma(s)ds)$ for $\forall t\in I$. Applying this to 
Eq.(\ref{eqn:C4}) by setting $a=0$, we readily have 
\begin{eqnarray}
D_{KL}(f_t\|f_\infty)\leqslant D_{KL}(f_0\|f_\infty)\exp(-\int^{t}_{0}\frac{2}{c}ds)=
e^{-\frac{2}{c}t}D_{KL}(f_0\|f_\infty),
\end{eqnarray}
where $f_0$ is the initial density function. This indicates explicitly that the distance between 
the initial and equilibrium density functions converges by the factor $e^{-2t/c}$. The smaller 
the value of $c$, the faster it converges.


\begin{thebibliography}{99}
\bibitem{Ffl} It is worth conferring that when we regard $\mathcal{P}$ as a function of $\beta$, 
then the Fisher information provides the fluctuation of energy for the canonical distribution: 
$I(\mathcal{P}(\beta))=\int \mathcal{P}(\beta) (d\ln\mathcal{P}(\beta)/d\beta)^2d\beta
=\langle H^2(\Gamma)\rangle_\mathcal{P}-\langle H(\Gamma)\rangle^2_\mathcal{P}$.
\end{thebibliography}
\end{document}